# 25-Fold Resolution Enhancement of X-ray Microscopy Using Multipixel Ghost Imaging

O. Sefi,[1,*] A. Ben Yehuda,[1] Y. Klein,[1,2] S. Bloch,[3] H. Schwartz,[1] E. Cohen[3] and S. Shwartz[1]

[1]*Physics Department and Institute of Nanotechnology and advanced Materials, Bar-Ilan University, Ramat Gan, 52900 Israel*
[2]*Nexus for Quantum Technologies, University of Ottawa, Ottawa, Ontario K1N 6N5, Canada.*
[3] *Faculty of Engineering and Institute of Nanotechnology and advanced Materials, Bar-Ilan University, Ramat Gan, 52900 Israel*
*\* sefior@biu.ac.il*

**Abstract:** Hard x-ray imaging is indispensable across diverse fields owing to its high penetrability. However, the resolution of traditional x-ray imaging modalities, such as computed tomography (CT) systems, is constrained by factors including beam properties, the absence of optical components, and detection resolution. As a result, typical resolution in commercial imaging systems is limited to a few hundred microns. This study advances high-photon-energy imaging by extending the concept of computational ghost imaging to multipixel ghost imaging with x-rays. We demonstrate a remarkable enhancement in resolution from 500 μm to approximately 20 μm for an image spanning 0.9 by 1 cm², comprised of 400,000 pixels and involving only 1000 realizations. Furthermore, we present a high-resolution CT reconstruction using our method, revealing enhanced visibility and resolution. Our achievement is facilitated by an innovative x-ray lithography technique and the computed tiling of images captured by each detector pixel. Importantly, this method can be scaled up for larger images without sacrificing the short measurement time, thereby opening intriguing possibilities for noninvasive high-resolution imaging of small features that are invisible with the present modalities.

## 1. Introduction

Hard x-rays offer unparalleled vision through opaque materials, enabling examination of elements invisible to longer wavelengths due to strong absorption. However, this advantage faces a hurdle in full-field imaging of large objects. The main challenge lies in the inherently small numerical aperture and reflectivity of optical components, which render them ineffective for the imaging of large objects [1]. Focusing and scanning with x-rays is possible through various techniques [1], but the scanning process is time-intensive due to the inherent tradeoff between resolution and field of view, hence scanning is impractical for large objects.

To circumvent these limitations, x-ray imaging systems employ cone beams that diverge and cover the large object. The resolution is mostly constrained by the resolution of the camera and by the size of the source [2]. On the detector side, the resolution is constrained by both the pixel size and by the thickness of the scintillation screen, responsible for converting x-rays into visible radiation. Regrettably, especially at high photon energies, thick scintillators are necessary to uphold a sufficiently high detection efficiency [3]. As a result, scattering within the scintillator introduces blurring in the detected image. The standard resolution in current x-ray imaging modalities is typically around a few hundred microns [2]. While this is suitable for various types of measurements, it hinders the ability to detect smaller features, such as microscopic cracks in materials or sub-millimeter subtle anomalies in organs. Current research endeavors aimed at resolution enhancement focus on the development of highly efficient scintillators, which can be made thinner, thus minimizing scattering effects [3]. Exploring alternative pathways for resolution enhancement, such as refining x-ray sources and incorporating better optical elements is an active field of research [4].

An alternative approach to achieve this goal involves the utilization of coded apertures or patterned illumination, employing nonuniformity to the beam to enhance image properties [5,6]. Ghost imaging (GI) is one such imaging modality, which utilizes intensity variations in the beam to retrieve the spatial information of an object. In classical GI a pair of identical nonuniform beams is used. One beam, known as the reference, is directed straight to a camera. The other beam, referred to as the test, illuminates an object and is captured by a single-pixel detector. This process is repeated multiple times with different light patterns. The image is then reconstructed by correlating the detected signals with the transmitted light patterns [7]. In its early days, GI was demonstrated with quantum light [8]. It was later extended to other types of radiation, such as thermal light, x-rays, atoms, neutrons, electrons, fluorescence, and scattered radiation [9–15]. Additionally, GI was extended to other domains, such as spectroscopy, diffraction imaging, polarization and temporal measurements, tomography, and object classification [16–21]. In one important variation of GI, called computational ghost imaging (CGI), the light patterns are introduced to the beam using a known mask, removing the need for their detection [22,23]. As the intensity patterns are known a single-pixel detector is sufficient for the image reconstruction.

One of the main challenges in GI lies in the number of intensity patterns needed, since it is proportional to the total number of pixels in the image. Despite the improved performance of current reconstruction algorithms in terms of compression ratio (the ratio between the number of reconstructed pixels to the number of measurements), a substantial number of light patterns are still necessary. The current approach leads to an extended measurement process, rendering previous methods for applying x-ray GI impractical for most applications. Significantly reducing the measurement time is a crucial step for advancing further developments in x-ray GI.

The second crucial challenge revolves around the modulation depth of light patterns. While in GI at optical wavelengths, achieving high-contrast intensity modulation is easily accomplished using Spatial Light Modulators, the absence of these components in the x-ray region poses a significant obstacle. In x-ray GI modulation is commonly achieved by directing x-rays through materials with a porous or grainy structure, or by using specially fabricated masks. To attain high resolution and contrast simultaneously, the features in the mask must be tall and narrow. However, the inherent instability and the small dimensions of these structures making them difficult to produce by standard techniques [24]. All prior efforts in x-ray GI have concentrated on photon energies below 30 keV. However, this limitation proves to be impractical for nearly all non-invasive x-ray measurements.

In this work, we demonstrate a method to overcome the two critical challenges described above by employing multi-pixel CGI and showcase its applicability for standard x-ray sources and flat panel detectors (FPDs). The high-resolution modulation of the x-ray beam is enabled by a custom-designed mask fabricated through advanced x-ray lithography. This mask, featuring incredibly tall and narrow features known as high aspect ratio structures, plays a crucial role in achieving the desired resolution. Unlike conventional GI, in this approach the number of measurements scales with the enhancement in resolution instead of the total number of pixels. This method, therefore, enables high-resolution reconstruction of image with a large field of view in a reasonable time frame.

## 2.   Experimental setup and image reconstruction

The experimental setup consists of a commercial tungsten x-ray tube that emits a cone beam with a peak photon energy of 80 keV, a shutter, a custom-designed gold mask, and an FPD (NDT0909M from iRay Technology Co. Ltd.) featuring a pixel pitch of 205 µm and a measured resolution of approximately 500 µm, as shown in Fig. 1(a). Additionally, we used a Zaber a X-RSW60A rotation stage to align the sample with respect to FPD, with accuracy of 0.07 degrees. The mask we used consists of rectangular $100 \pm 10$ µm tall gold pillars with a width of $15 \pm 1.5$ µm on a silicon substrate that is 525 µm thick. This height results in a modulation depth of approximately 50%, which is sufficient for our purpose. The mask spanned a total area of $1.4 \times 1.5$ cm². We designed the mask in segments, each containing a random matrix, while ensuring that different parts of the mask contain new information. The mask was fabricated by Microworks GmbH using state-of-the-art x-ray lithography. This technique allows for the fabrication of 3D structures with high accuracy and large aspect ratios, thanks to the deep penetration capabilities of the x-rays [25]. A high-resolution image of a segment from the mask is show in Fig. 1(b). The mask was positioned on two Thorlabs LTS150 translation stages (one horizontal and one vertical), with a precision of ±5 µm. The x-ray source was positioned approximately 1 m away from the mask, which was closely positioned to the object. We reconstructed the image by capturing low-resolution images of the object with the FPD for various positions of the mask, each corresponding to a different pre-designed illumination pattern.

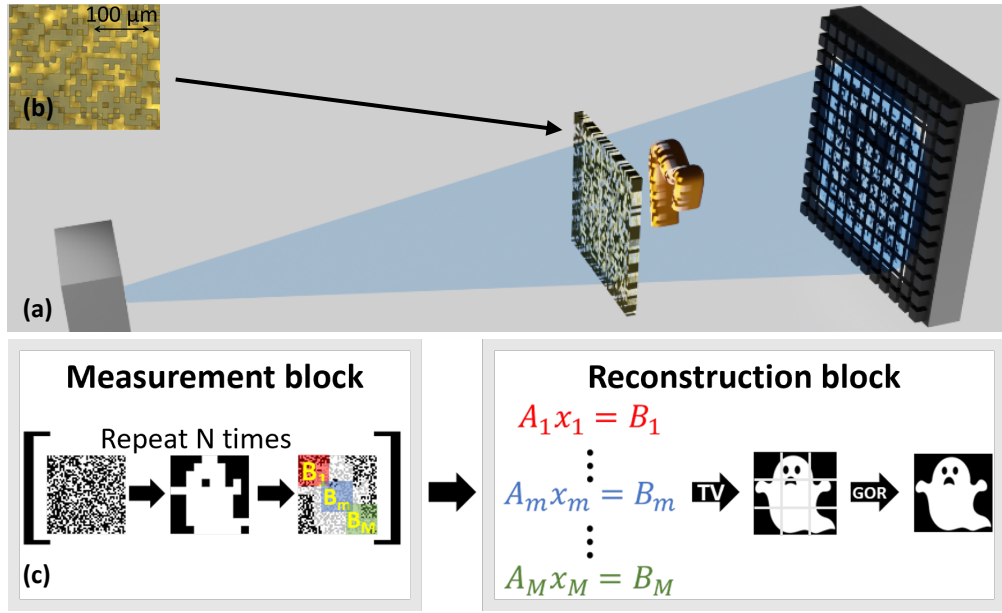

Figure 1 – (a) Schematic description of the experimental system. The cone beam propagates from left to right, impinges upon a mask that introduces inhomogeneous patterns on the object. The radiation transmitted through the object is collected by an FPD. (b) A magnified image of our mask. (c) Schematic description of the measurement and analysis process. N different light patterns illuminate the object and are collected by the FPD with each set of a few adjacent pixels serving as a bucket detector. The TV algorithm is used to solve the GI equations within each bucket area, resulting in a series of partially overlapping local images. Finally, the GOR algorithm is used to assemble all the tiles, forming the complete high-resolution image.

For each of the segments we defined, the image reconstruction was accomplished by solving the linear inverse problem [26]. This can be formulated as:

$$Ax - B = 0 \qquad (1).$$

Here, the illumination patterns are represented by the matrix A, in which every row is a single light pattern, the vector x represents the unknown transmission function of the object, and B is a vector representing the detected test signals. The resolution of x is determined by the feature size of the mask [17]. An exact solution of Eq. (1) requires the number of rows in A to be equal to the number of elements in x. Even for one segment from the object this can be time-consuming for high-resolution imaging (for instance, in our experiment we used 1000 measurements). However, the problem can be treated in several ways that approximate a solution with a reduced number of light patterns.

We used compressive sensing (CS) [27,28], which is widely adopted for solving such inverse problems. This approach leverages prior knowledge about the spatial characteristics of the object, such as sparsity, to accelerate the process of the reconstruction [27]. CS introduces constraints (or regularization terms) to Eq. (1) and attempts to minimize the resulting expression. In this work, we used the total variation (TV) as our constraint. The TV problem can be formulated as the minimization task of:

$$TV = \mu \|Ax - B\|_2 + \lambda_1 \|\nabla_x x\|_1 + \lambda_2 \|\nabla_y x\|_1 \qquad (2)$$

where $\|\cdots\|_2$ and $\|\cdots\|_1$ are the $L^2$ and $L^1$ norms, respectively, and $\nabla_x$ and $\nabla_y$ represent the first derivative in the x and y directions. The parameters μ and λ serve as regularization factors, which balance between the actual measurements, as given by Eq. (1), and the smoothing constraints captured by the spatial derivatives. This approach considerably reduces the number of required light patterns. However, both the required number of light patterns and the measurement time scale with the number of pixels in the image, previously restricting the application of CGI in large images.

A schematic description of the reconstruction workflow is shown in Fig 1(c). First, we applied the TV reconstruction algorithm to each set of few adjacent pixels on the FPD, which results in a series of partially overlapping local images. Finally, we assembled all the tiles, using globally optimized registration (GOR), to create a complete high-resolution image with a large field of view [29].

## 3. RESULTS

To demonstrate the advantages of our approach in reconstructing large images at enhanced resolution, we imaged a copper wire with a minimum thickness of ~130 μm and total area of about 1×1 cm$^2$. We binned each 5×5 pixel block of the FPD to form an array of partially overlapping bucket detectors. The chosen block size is optimal for our system, striking a balance that ensures a reasonably low compression ratio while minimizing undesirable effects arising from the boundaries. These effects may result from sub-FPD-pixel misalignment of the mask or the nonuniform response of the FPD within the pixel size. The low-resolution FPD image and the high-resolution reconstruction are shown in Fig. 2 (a) and (b), respectively. The superior resolution of our method is evident. Previously obscured features like thin edges of the copper wire are now clearly resolved and exhibit high contrast in the reconstructed image, as further supported by the cross-sections of Fig. 2 (c) and (d). In these cross-sections the green and the red dots represent the data from the FPD, and the purple and blue dots represent the data from our reconstruction. More importantly, the limitations of the FPD image with its low resolution are overcome by our method, which unveils intricate details significantly smaller than the FPD pixel size.

To quantitatively assess the resolution of our reconstruction, we examined the blurring of a sharp edge of a 0.7 mm thick polished silicon wafer, with blurring of about 15 μm due to the resolution of the motor. The FPD image of the silicon wafer and a high-resolution reconstruction are shown in Fig. 3 (a) and (b). The cross-sections are depicted in Fig. 3 (c), where the green and red dots correspond to measurements by the FPD, and our reconstruction, respectively.

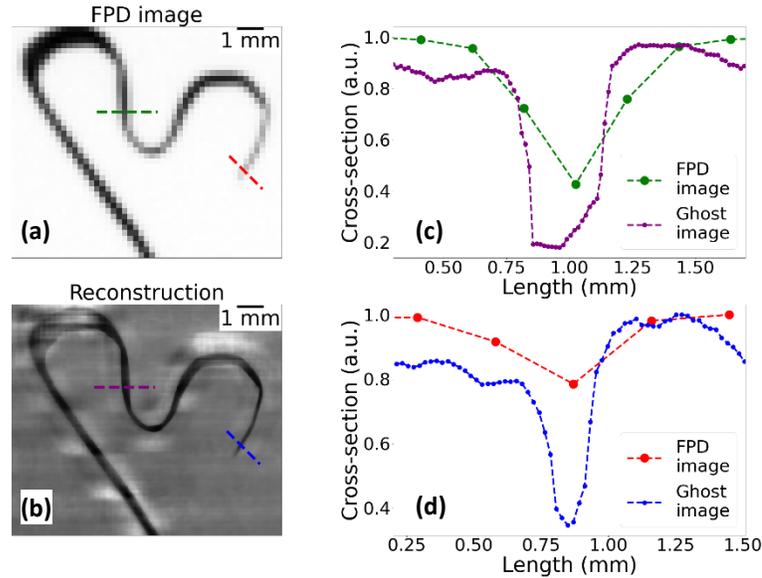

Figure 2 – (a) The image of a copper wire captured with the FPD (b) The high-resolution reconstruction with the multipixel GI approach. (c) and (d) Cross-sections along the dashed lines with the corresponding colors in the images. In these cross-sections, the green and the red dots represent the FPD, and the purple and the blue dots represent the reconstruction. The dashed lines serve as guides to the eye.

By measuring the half-width of the slope of the cross-section near the edge, we estimate the resolution to be approximately 20 μm. This result is consistent with the feature size of the mask. Since the resolution of the FPD is nearly 500 μm, we conclude that our approach enhances the resolution by a factor of 25.

Furthermore, even though the resulting image in Fig. 2b comprises approximately 250,000 pixels, it was reconstructed using only 1000 realizations. Given that the measurement time is independent of the number of segments, this implies that our method is capable of reconstructing images with a pixel count commonly found in noninvasive imaging, all within acceptable measurement times.

We acknowledge that, in principle, the number of realizations can be further reduced based on the specific object and the prior knowledge about it, and improvements can be achieved through advanced reconstruction algorithms. To assess the impact of the number of realizations on the quality of the reconstructed image, we deliberately chose a smaller subset from our realizations. We conducted multiple reconstructions of a segment from the copper wire, each time utilizing different realizations from this selected subset. The goal of this process is to obtain multiple reconstructions under similar imaging conditions, from which we can derive statistical data on the reconstruction quality. We compute the mean squared error (MSE) and its standard deviation in relation to the optimal reconstruction generated from 1000 realizations. We continue this process with subsets of increasing size to demonstrate the dependency of the reconstruction on the number of realizations. The results are shown in Fig 3 (d). Here, the blue dots represent the MSE from the best image and the vertical error-bars represent its standard deviations. In the insets the reconstructions with 20, 150, 500 and 1000 realizations are shown form left to right, respectively. It is evident that the significant improvement in the image occurs after about 500 realizations, which corresponds to a compression ratio of about 10 within each

5x5 pixel block of the FPD. This result can be further improved by using better reconstruction algorithm [30–32].

The results presented above were obtained using a bucket size of 5×5 pixels. However, since we use a multi-pixel detector for the multi-pixel GI, a question raises: is there an optimal bucket size that optimizes the quality of the images for a chosen number of realizations? This is because we aim to reduce the number of realizations to reduce the measurement time, but in GI the quality of image improves with the number of realizations. This consideration suggests that the best choice would be one pixel of the FPD. However, due to scattering from the scintillation screen of the FPD the point spread function of the detector is larger than one pixel, which suggests that selecting one pixel is not the optimal choice. In addition, our reconstruction algorithms also strongly depend on the size of the pixel due to the degree of sparsity and overlap issues and small bucket sizes are more susceptible to errors in the sub-FPD-pixel-size positioning of the mask. To address this question, we show the dependence of the reconstructed image on the bucket size in Fig. 3 (e). We calculated the acutance of the reconstructed silicon edge, which is the mean value of the image gradient, approximated by the Sobel filter [33–35]. We repeated this for different bucket sizes ranging from 2×2 to 10×10 FPD pixels. We reiterated the validation process mentioned above for each bucket size by using 1000 different realizations. In the Fig. 3 (e), the blue dots correspond to the acutance, and the vertical error-bars represent its standard deviations. The insets show the reconstructed images for buckets sizes of 2×2, 5×5 and 10×10 FPD pixels, from left to right, respectively. We concluded that for our system, the optimum is for buckets of 5×5 FPD pixels, a number that may vary for different systems and different reconstruction algorithms.

To further demonstrate the power of our method, we performed a high-resolution computational tomography (CT) reconstruction of a M4 screw with a washer and half a nut. The 3D reconstructed volume has dimensions of about $8.6 \times 8.6 \times 8.8$ mm$^3$ (L×W×H) and contains approximately $2 \cdot 10^8$ voxels. We used 73 projections, each with 1089 realizations, for the reconstruction. The CT reconstruction was performed using the TIGRE toolbox [36]. Fig. 4 presents the tomograms, with the top row showing reconstructions from the FPD images and the bottom row displaying those from our high-resolution images. Specifically, panels (a) and (b) depict a top-view slice of the bisected nut, panels (c) and (d) illustrate a front-view slice, and panels (e) and (f) reveal a side-view slice. More tomograms are shown in Fig. S1 in the supplementary materials. Clearly, our method significantly enhances the visibility of fine details, such as the hexagonal shape of the nut and the pitch of the screw, which are barely discernible in the FPD CT reconstruction.

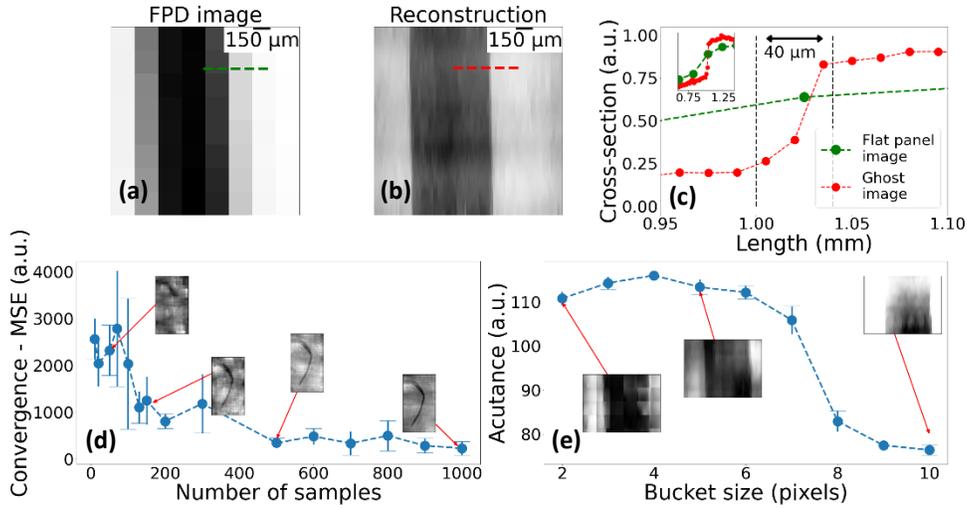

Figure 3 – (a) FPD image of the silicon wafer. (b) high-resolution reconstruction with the multipixel GI approach. (c) Cross-section along the dashed lines. The green dots represent the FPD, and the red dots represent the reconstruction. In the inset, the zoomed-out cross-section is shown. (d) MSE of the copper wire images dependence on the number of realizations. The blue dots represent the MSE, and the vertical error-bars represent its standard deviation. Insets (left to right): reconstructions with 20, 150, 500, and 1000 realizations. (e) Acutance of the reconstructed silicon wafer image vs. bucket size. Blue dots represent the acutance and vertical error-bars represent its standard deviation. Insets (left to right): Reconstructions with bucket sizes of 2, 5 and 10 FPD pixels. Dashed lines serve as guides to the eye.

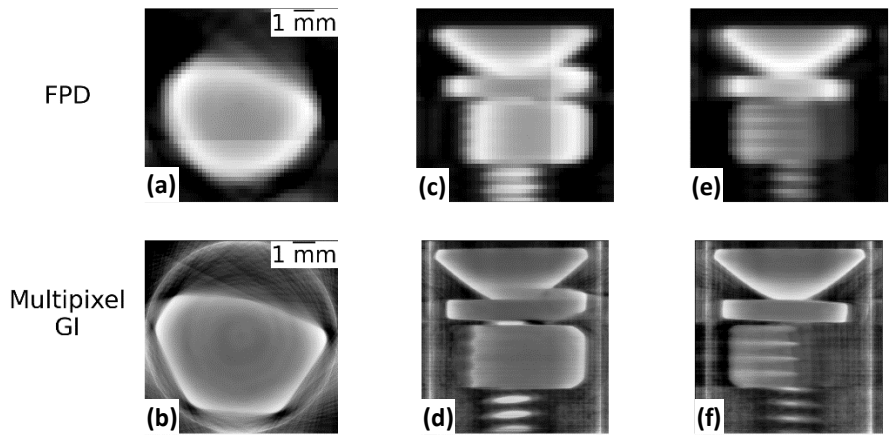

Figure 4 – Comparison between tomograms of a M4 screw reconstructed using the FPD (top row) and using our multipixel GI method (bottom row). The tomograms are shown from top view (a and b), from front view (c and d) and from side view (e and f).

In conclusion, we have introduced and demonstrated a cutting-edge x-ray imaging technique operating at high photon energies, leveraging patterned light and intensity correlations. Our method successfully reconstructed high-resolution 2-dimensional images and a 3-dimensional volume, with the former comprising 400,000 pixels reconstructed with 1000 realizations. This achievement translates to a remarkable 100-fold reduction in measurement time compared to common CGI, maintaining equivalent object size and resolution. This achievement sums up to a reduction of about $7 \cdot 10^4$ in measurement time for the high-resolution CT. Further enhancements in efficiency can be pursued through the integration of additional prior knowledge into reconstruction algorithms or by incorporating advanced deep learning techniques.

Notably, our approach demonstrates scalability for larger images with a comparable number of realizations, achievable through mask size expansion using contemporary technologies. Future advancements may focus on achieving even higher resolutions by employing masks with smaller feature sizes. Our straightforward setup, showcasing robustness and high-quality image reconstruction, unlocks unprecedented capabilities for high resolution imaging and CT, at high photon energies. This enables the visualization of minute features currently beyond the reach of existing methods. Examples for applications that would particularly benefit from our modality include high-resolution non-destructive imaging and mammography.

**References**

1. A. Guilherme, G. Buzanich, and M. L. Carvalho, "Focusing systems for the generation of X-ray micro beam: An overview," Spectrochim Acta Part B At Spectrosc 77, 1–8 (2012).

2. W. Huda and R. B. Abrahams, "X-Ray-Based Medical Imaging and Resolution," American Journal of Roentgenology 204, W393–W397 (2015).
3. L. Wollesen, F. Riva, P.-A. Douissard, et al., "Scintillating thin film design for ultimate high resolution X-ray imaging," J Mater Chem C Mater 10, 9257–9265 (2022).
4. A. Sakdinawat and D. Attwood, "Nanoscale X-ray imaging," Nat Photonics 4, 840–848 (2010).
5. E. Mojica, C. V. Correa, and H. Arguello, "High-resolution coded aperture optimization for super-resolved compressive x-ray cone-beam computed tomography," Appl Opt 60, 959 (2021).
6. J. Li, S. Chen, D. Ratner, et al., "Nanoscale chemical imaging with structured X-ray illumination," Proceedings of the National Academy of Sciences 120, (2023).
7. Y. Bromberg, O. Katz, and Y. Silberberg, "Ghost imaging with a single detector," Phys Rev A (Coll Park) 79, 053840 (2009).
8. T. B. Pittman, Y. H. Shih, D. V. Strekalov, et al., "Optical imaging by means of two-photon quantum entanglement," Phys Rev A (Coll Park) 52, R3429–R3432 (1995).
9. A. Valencia, G. Scarcelli, M. D'Angelo, et al., "Two-Photon Imaging with Thermal Light," Phys Rev Lett 94, 063601 (2005).
10. D. Pelliccia, A. Rack, M. Scheel, et al., "Experimental X-Ray Ghost Imaging," Phys Rev Lett 117, 113902 (2016).
11. R. I. Khakimov, B. M. Henson, D. K. Shin, et al., "Ghost imaging with atoms," Nature 540, 100–103 (2016).
12. A. M. Kingston, G. R. Myers, D. Pelliccia, et al., "Neutron ghost imaging," Phys Rev A (Coll Park) 101, 053844 (2020).
13. S. Li, F. Cropp, K. Kabra, et al., "Electron Ghost Imaging," Phys Rev Lett 121, 114801 (2018).
14. Y. Klein, O. Sefi, H. Schwartz, et al., "Chemical element mapping by x-ray computational ghost fluorescence," Optica 9, 63 (2022).
15. A. Ben Yehuda, O. Sefi, Y. Klein, et al., "High-resolution computed tomography with scattered x-ray radiation and a single pixel detector," Comms. Eng (accepted).
16. Y. Klein, E. Strizhevsky, F. Capotondi, et al., "High-resolution absorption measurements with free-electron lasers using ghost spectroscopy," in Frontiers in Optics + Laser Science 2022 (FIO, LS) (Optica Publishing Group, 2022), p. LW7F.4.
17. F. Ferri, D. Magatti, A. Gatti, et al., "High-Resolution Ghost Image and Ghost Diffraction Experiments with Thermal Light," Phys Rev Lett 94, 183602 (2005).
18. P. Janassek, S. Blumenstein, and W. Elsäßer, "Recovering a hidden polarization by ghost polarimetry," Opt Lett 43, 883 (2018).
19. P. Ryczkowski, M. Barbier, A. T. Friberg, et al., "Ghost imaging in the time domain," Nat Photonics 10, 167–170 (2016).
20. O. Sefi, Y. Klein, E. Strizhevsky, et al., "X-ray imaging of fast dynamics with single-pixel detector," Opt Express 28, 24568 (2020).
21. J.-N. Cao, Y.-H. Zuo, H.-H. Wang, et al., "Single-pixel neural network object classification of sub-Nyquist ghost imaging," Appl Opt 60, 9180 (2021).
22. J. H. Shapiro, "Computational ghost imaging," Phys Rev A 78, 061802 (2008).
23. Y. Klein, A. Schori, I. P. Dolbnya, et al., "X-ray computational ghost imaging with single-pixel detector," Opt Express 27, 3284 (2019).
24. T. Mappes, S. Achenbach, and J. Mohr, "X-ray lithography for devices with high aspect ratio polymer submicron structures," Microelectron Eng 84, 1235–1239 (2007).
25. A. Bharti, A. Turchet, and B. Marmiroli, "X-Ray Lithography for Nanofabrication: Is There a Future?," Frontiers in Nanotechnology 4, (2022).
26. A. Ribes and F. Schmitt, "Linear inverse problems in imaging," IEEE Signal Process Mag 25, 84–99 (2008).
27. O. Katz, Y. Bromberg, and Y. Silberberg, "Compressive ghost imaging," Appl Phys Lett 95, 131110 (2009).
28. Y. Sher, "Review of Algorithms for Compressive Sensing of Images," arXiv prepr. arXiv:1908.01642 (2019).
29. S. Preibisch, S. Saalfeld, and P. Tomancak, "Globally optimal stitching of tiled 3D microscopic image acquisitions," Bioinformatics 25, 1463–1465 (2009).
30. S. Liu, X. Meng, Y. Yin, et al., "Computational ghost imaging based on an untrained neural network," Opt Lasers Eng 147, 106744 (2021).
31. H. Wu, G. Zhao, M. Chen, et al., "Hybrid neural network-based adaptive computational ghost imaging," Opt Lasers Eng 140, 106529 (2021).
32. F. Wang, C. Wang, M. Chen, et al., "Far-field super-resolution ghost imaging with a deep neural network constraint," Light Sci Appl 11, 1 (2022).
33. R. M. Rangayyan, N. M. El-Faramawy, J. E. L. Desautels, et al., "Measures of acutance and shape for classification of breast tumors," IEEE Trans Med Imaging 16, 799–810 (1997).
34. R. Ranjan and V. Avasthi, "Edge Detection Using Guided Sobel Image Filtering," Wirel Pers Commun 132, 651–677 (2023).
35. R. O. Duda, P. E. Hart, and D. G. Stork, Pattern Classification (2000).
36. A. Biguri, M. Dosanjh, S. Hancock, et al., "TIGRE: a MATLAB-GPU toolbox for CBCT image reconstruction," Biomed Phys Eng Express 2, 055010 (2016).

# 25-FOLD RESOLUTION ENHANCEMENT OF X-RAY MICROSCOPY USING MULTIPIXEL GHOST IMAGING: SUPPLEMENTAL DOCUMENT

## 1. Methods

To compute the acutance of our reconstructed images, we applied the Sobel filter with kernel size of 5 on the image. The Sobel kernels $G_x$ and $G_y$ are given by [1]:

$$G_x = \begin{pmatrix} -1 & -2 & 0 & 2 & 1 \\ -4 & -8 & 0 & 8 & 4 \\ -6 & -12 & 0 & 12 & 6 \\ -4 & -8 & 0 & 8 & 4 \\ -1 & -2 & 0 & 2 & 1 \end{pmatrix}; \quad G_y = \begin{pmatrix} -1 & -4 & -6 & -4 & -1 \\ -2 & -8 & -12 & -8 & -2 \\ 0 & 0 & 0 & 0 & 0 \\ 2 & 8 & 12 & 8 & 2 \\ 1 & 4 & 6 & 4 & 1 \end{pmatrix} \quad \text{(S1)}$$

for a given image A, the imaged is convolved with the Sobel kernels and the resulting filtered image G is:

$$G = \sqrt{(G_x \cdot A)^2 + (G_y \cdot A)^2} \quad \text{(S2)}$$

## 2. Additional results

In Fig. S1 we provide additional tomograms of the M4 screw. The high-resolution reconstruction and the FPD reconstruction are shown in the left and right panels, respectively. The top two rows contain tomograms shown from front view while the two rows in the bottom show tomograms from side view.

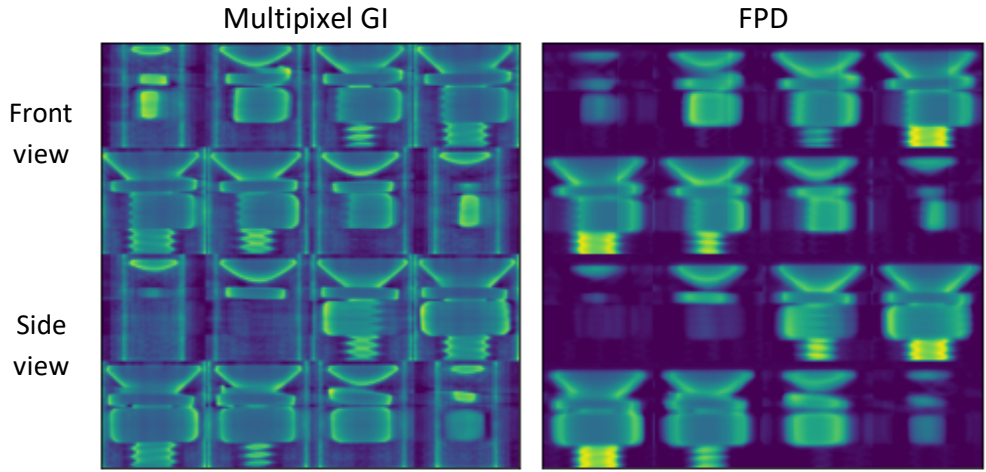

Fig. S1. –Tomograms of a M4 screw reconstructed using our multipixel GI method (left panel) and the FPD (right panel). The top two rows are shown from front view and the two rows at the bottom are shown from side view.

Additional visualization of the high-resolution reconstruction is provided in the attached video files.

## References

37. R. O. Duda, P. E. Hart, and D. G. Stork, *Pattern Classification* (2000).